\documentclass[preprint,showpacs,preprintnumbers,amsmath,amssymb]{revtex4}
\usepackage{graphicx,amsmath,amssymb,amsfonts,latexsym,color,dcolumn,bm,epsfig,subfigure}

\begin{document}

\title{Electrostatic Background Forces due to Varying Contact Potentials in Casimir Experiments}

\author{Steve K. Lamoreaux}
\affiliation{Yale University, Department of Physics,
P.O. Box 208120, New Haven, CT 06520-8120, USA}

\date{\today}

\begin{abstract}

The existence of a monotonic distance dependent contact potential between two plates in a Casimir experiment leads to an additional electrostatic force that is significantly different from the case of a constant potential.  Such a varying potential can arise if there is a uniform gradient in the work function or contact potential across a plate, as opposed to random microscopic fluctuations associated with patch potentials.  A procedure to compensate for this force is described for the case of an experiment where the electrostatic force is minimized at each measurement distance by applying a voltage between the plates. It is noted that the minimizing voltage is not the contact potential.

\end{abstract}


\maketitle

In our recent work to measure the short range attractive force between a spherical and flat pure Ge plates, two ``spurious" have been observed.  In this experiment, the force is measured as a function of plate separation, and as a function of voltage applied between the plates at a specific separation.  The force at the voltage which minimizes the force at each separation was thought to represent the pure ``Casimir" force between the plates. However, the applied voltage $V_a(d)$ required to minimize the (electrostatic) force is observed to depend on $d$, and is of the form (in the 1-50 $\mu$m range)
\begin{equation}
V_a(d)=a\log d +b
\end{equation}
where $a$ and $b$ are constants with magnitude of a few mV.  It appears that this variation is not due to simple patch effects, but more likely due to a slight variation in contact potential (or work function) across the faces of the plates.  This specific form arises if it is assumed that the voltage on the surface at a radial distance $r$ from the center of a plate varies as $r^n$ where $n<<1$, and a suitable average electrostatic force determined as a function of $d$. Of course the simple log form breaks down at large separations where the plates look like two disks, at at short distances, corresponding to the characteristic Debye length $\lambda/\epsilon$, where the potential will also tend to become constant.\cite{sklarxiv}

The second spurious effect is the appearance of a long-range $1/d$-like potential for the minimized force.  An analysis suggests that this force is better described as $1/d^m$ where $m\approx 1.2-1.4$.  

As we show here, the variation in $V_a(d)$ implies an attractive force that increases as $1/d^{1.25}$. 

An understanding of the specific origin of the variation of $V_a(d)$ is not necessary to correct for the additional force that it causes, we simply need the experimentally determined $V_a(d)$.

We note further that $V_a(d)$ is not a measure of the contact potential, but the voltage which minimizes the force.

In performing the experiment, at each separation $d$, $V_a$ is varied and its value that minimizes the attractive force is determined.
Specifically, the electrostatic energy contained in the field between (and within, due to finite Debye length) the plates is given by 
\begin{equation}
{\cal E}(d)={1\over 2} C(d) (V_a+V_c(d))^2
\end{equation}
where $C(d)$ is the capacitance between the plates, $V_a$ is the applied potential and is an independent variable, and $V_c(d)$ is the average weighted contact potential between the plates.

The force between the plates is given by the derivative of $\cal E$,
\begin{equation}
F(d)={\partial {\cal E}(d)\over \partial d}={1\over 2} {\partial C(d)\over\partial d}(V_a+V_c(d))^2+C(d)(V_a+V_c(d)){\partial V_c(d)\over \partial d}
\end{equation} 

Now the minimum in the force is determined by the derivative with $V_a$:
\begin{equation}
{\partial F(d)\over \partial V_a}={\partial C(d)\over\partial d}(V_a+V_c(d))+C(d){\partial V_c(d)\over \partial d}=0
\end{equation}
which determines $V_a(d)$.  Thus,
\begin{equation}
{\partial V_c(d)\over \partial d}=-{1\over C(d)}{\partial C(d)\over\partial d}(V_a(d)+V_c(d))
\end{equation}
which allows the determination of $V_c(d)$ when $V_a(d)$ is known.  

The differential equation can be solved numerically, noting that at long distances $V_a(d)=-V_c(d)$, and that $V_c(d)$ become constant.

The electrostatic force between the plates at the minimized potential is given by
\begin{equation}
F(d)=\left({\partial E(d)\over \partial d}\right)_{V_a=const}={1\over 2} \left({\partial [C(d)(V_a(d)+V_c(d))^2]\over \partial d}\right)_{V_a=const}.
\end{equation}
Although $V_a(d)$ is determined, it is externally set and does not vary directly with the plates' motion, hence its derivative should not contribute to the force.
To calculate the force between the plates,
this derivative is evaluated numerically using the measured $V_a(d)$ and inferred $V_c(d)$.  $C(d)$ is determined from the parallel plate (Debye screening corrected for Ge \cite{sklarxiv}) capacitance, by use of the pairwise additive approximation.  The derivative $C'(d)$ is determined numerically.  $C(d)$ evaluated in this fashion agrees extremely well with direct measurement of $C(d)$.

So far, this procedure appears to describes our observed extra force with $1/d$ character.  However it should be noted that $V_a(d)$ needs to be measured at very large separations in order to set the numerical integration initial condition, approximately 100 times the distance of closest approach.  If measurements to this distance are performed, this extra electrostatic force can be accounted for with no adjustable parameters. 

There are some nice features to this result.  First, if we apply a constant offset $V_0$ to $V_c(d)$, this effect is compensated by $V_a(d)-V_0$ which is easily seen as the relationship is linear.  

Second, if we assume $V_a(d)=\log(d)$, and take $C(d)=-\log(d)$ (up to multiplicative constants for both), we obtain 
\begin{equation}
F(d)\propto {(\log(d))^2\over d}
\end{equation}
which has a form of $1/d^{1.25}$ over any  small range of $d<<1$.

In the case where $V_a$ is not adjusted at each measurement point to the minimizing value, the force can be directly determined from Eq. (3) above.

It should be emphasized that any precision measurement of the Casimir force requires verification that the contact potential is not changing as a function of distance, and if it is, a correction to the force is required using a procedure similar to that outlined here.


\begin{thebibliography}{100}

\bibitem{sklarxiv} S.K. Lamoreaux, arXiv:0801.1283.

\end{thebibliography}
\end{document}